\documentclass[universe,article,submit,moreauthors,pdftex,10pt,a4paper]{mdpi} 
\let\originalleft\left
\let\originalright\right
\renewcommand{\left}{\mathopen{}\mathclose\bgroup\originalleft}
\renewcommand{\right}{\aftergroup\egroup\originalright}

\newcommand{\fracd}[2]{\frac{\displaystyle\strut {#1}}{\displaystyle\strut {#2}}}
\newcommand{\recd}[1]{\frac{\displaystyle 1}{\displaystyle\strut {#1}}}
\renewcommand{\r}[1]{(\ref{#1})}
\newcommand{\z}[1]{\left({#1}\right)}

\newcommand{\sz}[1]{\left[{#1}\right]}

\newcommand{\m}[1]{\mathrm{#1}}
\renewcommand{\v}[1]{\mathbf{#1}}

\newcommand{\gvec}[1]{\mbox{\boldmath${#1}$}}
\firstpage{1} 
\articlenumber{x}
\doinum{10.3390/------}
\pubvolume{xx}
\pubyear{2018}
\copyrightyear{2018}
\history{Received: date; Accepted: date; Published: date}

\pdfoutput=1

\nolinenumbers
\Title{
Scaling properties of spectra in new exact solutions of rotating, multi-component fireball hydrodynamics
}

\Author{Tam\'as Cs\"org\H{o}$^{1,2}$, and G\'abor Kasza $^{2}$}

\AuthorNames{Tam\'as Cs\"org\H{o} and G\'abor Kasza}

\address{%
$^{1}$ Wigner RCP, H - 1525 Budapest 114, P.O.Box 49, Hungary\\
$^{2}$ EKU KRC, H-3200 Gy\"ongy\"os, M\'atrai \'ut 36, Hungary}

\abstract{
We describe fireballs that rehadronize from a perfect fluid of quark matter,
characterized by the lattice QCD equation of state, to a chemically frozen,
multi-component mixture, that contains various kinds of observable hadrons.
For simplicity and clarity, we apply a non-relativistic approximation to
describe the kinematics of this expansion. Unexpectedly, we identify a
secondary explosion that may characterize fireball hydrodynamics at the QCD
critical point.  After rehadronization, the multi-component mixture of hadrons
keeps on rotating and expanding together, similarly to a single component
fluid.  After kinetic freeze-out, the effective temperature $T_{i}$ of the
single-particle spectra of hadron type $h_i$  is found to be a sum of the
kinetic freeze-out temperature $T_f$ (that is independent of the hadron type
$h_i$) and  a term proportional to the mass $m_i$ of hadron type $h_i$.  The
coefficient of proportionality to $m_i$ is also found to be independent of the
hadron type $h_i$ but be dependent on the radial flow and vorticity of
collective dynamics.
}


\usepackage{subcaption}
\usepackage{chngcntr}
\preto{\abstractkeywords}{\nolinenumbers}

\begin{document}

\nolinenumbers



\section{Introduction}

In this paper, the main theme of the analysis is the mass systematics of the
single particle spectra as measured in relativistic heavy ion collisions, based
on exact solutions of fireball hydrodynamics. Interestingly, the NA44
Collaboration observed signals of collective expansion in symmetric heavy-ion
collisions in 200 AGeV S+S and 158 AGeV Pb+Pb reactions at CERN SPS: the
transverse mass spectra of pions, kaons and protons was found to be nearly
exponential in these reactions, with a slope parameter that increased nearly
linearly with the mass of the observed hadrons~\cite{Bearden:1996dd}.  This
linear mass dependence of the slope parameters was interpreted as an evidence of
collective transverse flow  in these heavy ion induced central collisions.

\begin{figure}
	\includegraphics[scale=0.5]{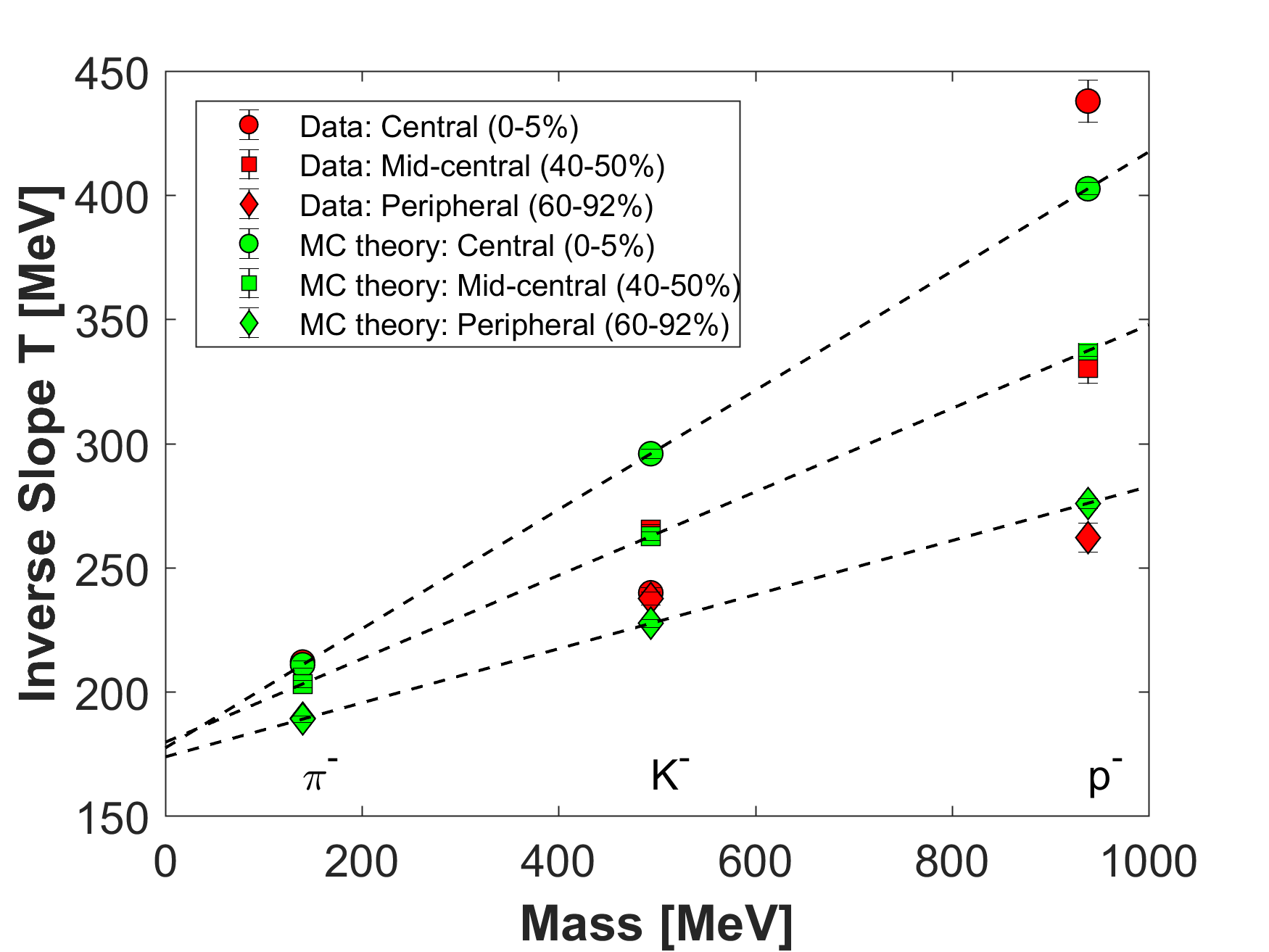}
	\includegraphics[scale=0.5]{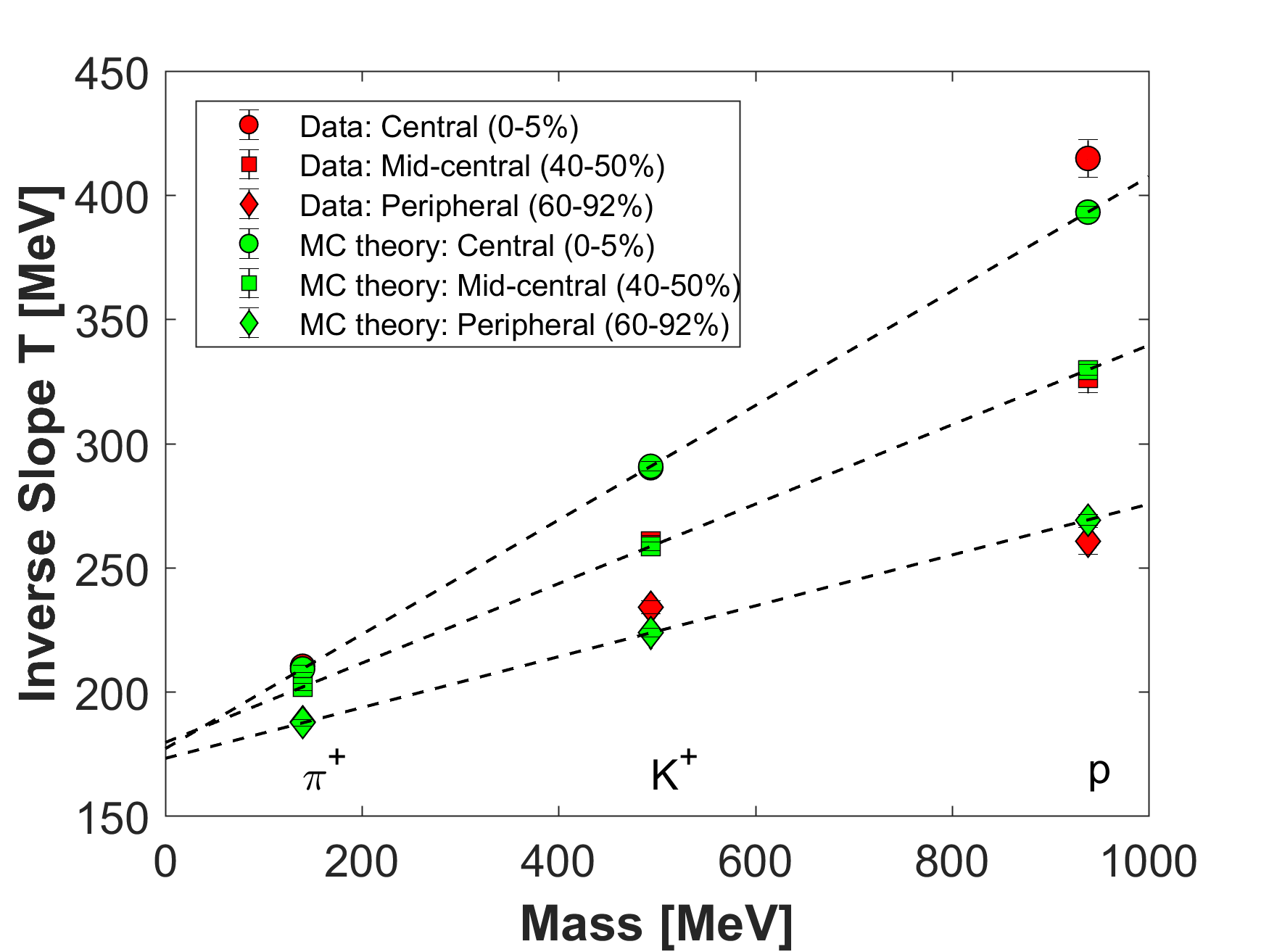}
	\centering
	\caption{
The mass and centrality dependence of the inverse slope parameters of
positively and negatively charged pions and kaons, as well as for protons and
anti-protons in $\sqrt{s}=$ 200 GeV Au+Au collisions, as measured in various
centrality classes by the PHENIX collaboration, based on Figure 10 and Table IV
of ref.~\cite{Adler:2003cb}. Dashed lines are drawn to guide the eye,
corresponding to eq.~(\ref{e:teffm}) and also to earlier theoretical results
for a single-component hadron gas in non-central heavy ion collisions as
detailed in refs.~\cite{Akkelin:2000ex,Csorgo:2001xm}. Green symbols indicate
our new result, based on our new exact solutions of fireball hydrodynamics for
a collectively expanding and rotating, multi-component hadronic matter, 
as indicated by symbols with ``MC theory" in the legends. These
green symbols appear at discrete values of hadronic masses, corresponding to eq.~(\ref{e:teffmi}),
 along the eye-guiding dashed lines.
}
	\label{fig:inverse_slope}	
\end{figure}

The inverse slope parameters for positively and negatively charged pions and
kaons,  as well as for protons and anti-protons has subsequently also been
measured in various heavy ion collisions, for example in $\sqrt{s_{NN}} = 200 $
GeV Au+Au collisions by the PHENIX Collaboration in three centrality bins,  as
is shown very clearly on Figure 10 and in Table IV  of
ref.~\cite{Adler:2003cb}, reproduced and detailed on
Fig.~\ref{fig:inverse_slope}.  These inverse slope parameters  were also found
to increase with increasing particle mass, in all centrality bins.  This
increase for central collisions was found to be more rapid for heavier
particles, taken as another indication of collective expansion dynamics not
only in central but also in peripheral heavy ion reactions at RHIC.  Such a
behavior was predicted -- under certain conditions corresponding to a
non-relativistic radial expansion dynamics -- in
refs.~\cite{Schnedermann:1993ws,Csorgo:1994fg,Csorgo:1995bi,Csizmadia:1998ef},
for central collisions. A similar behaviour was derived for  non-central heavy
ion collisions as well,  in refs.~\cite{Akkelin:2000ex,Csorgo:2001xm}.

The early derivations were based on the blast-wave~\cite{Schnedermann:1993ws},
or on the Buda-Lund model~\cite{Csorgo:1994fg,Csorgo:1995bi} that focused on a
parameterization of the particle emission in phase-space around the time of
kinetic freeze-out.  It is remarkable that the transverse, radial flows in both
the blast-wave and the Buda-Lund parameterization go back to precisely the same
exact solution of non-relativistic fireball hydrodynamics, namely the
Zim\'anyi-Garpman-Bondorf solution of ref.~\cite{Bondorf:1978kz}. A few years
after the successfull hydrodynamical parameterizations of the freeze-out
phase-space distribution were obtained, the time evolution was also accessed
with the help of the first exact solution within the Buda-Lund family of exact
solutions of fireball hydrodynamics. The first of such Buda-Lund hydro
solutions was found for a spherically symmetric, radially expanding fireball
with spatially homogeneous temperature profile and a Gaussian density profile
in ref.~\cite{Csizmadia:1998ef}, where the  inverse slope parameter of the
single particle spectra, $T_{\mbox{\small\it eff}}$ has also been shown to scale
as an affine-linear function of the particle mass $m$:
\begin{equation}
T_{\mbox{\small\it eff}}(m)
= T_f + m\langle u_T \rangle^2, \label{e:teffm}
\end{equation}
where $m$ is the mass of a single kind of
elementary particle with three kinetic degree of freedom, that consists the
expanding medium.  The average transverse flow is denoted by
$\langle u_T \rangle$ and the kinetic freeze-out temperature is denoted by $T_f$. Although the above relation is typically quoted as a justification for the linear mass dependence of the effective
temperature in a hot and dense, hadronic matter, actually the derivations
had a much more limited scope, evaluating the mass dependence for one kind of
hadrons only, while in the experiment,  a mixture of various hadronic components is
observed.

Such an experimental result is beautifully illustrated on Fig. 10.  of
ref.~\cite{Adler:2003cb} and reproduced also on our
Fig.~\ref{fig:inverse_slope}, that compares measured slope parameters in
relativistic heavy ion collisions at mid-rapidity with eq.~\eqref{e:teffm},
derived in a non-relativistic context. The success of this comparision suggests
that the key scaling properties of the transverse mass spectra at mid-rapidity
may perhaps be understood in the framework of non-relativistic kinematics.
Inspired by this insight, we try to handle the equations of relativistic and
non-relativistic hydrodynamics in a similar way in Section~\ref{s:hydro}. 
For the sake of clarity and brevity, we present new, rotating ellipsoidal exact solutions of
fireball hydrodynamics only in the non-relativistic limit,  as detailed in
Section~\ref{s:rotsol}.

Note, however, that all the earlier derivations dealt with a hydrodynamically
expanding medium that had only a single component, i.e. only one kind of
hadrons, with a given mass $m$. The linear rise in the data was compared with
the results of the calculations by extrapolating the theoretical results as a
continuous function of the mass $m$, that was considered as a smoothly varying
parameter of the solution and the resulting effective temperature or slope
parameter.  However, the experimentally observed mass spectrum of the hadrons
is essentially discrete, and not a smoothly varying function. In the data
analysis, for example, the slope parameters were measured at  the mass of
pions, kaons and (anti)protons, as indicated on Fig.~\ref{fig:inverse_slope}.
One of the challenges considered in this work is to generalize the derivation
from a single-component hadron gas to a multi-component hadron gas, that is  a
mixture of hadrons with different discrete values of their masses. Such a
derivation  seemed to be an almost hopeless theoretical challenge, as far as we
know it was not even attempted before.  This challenge however is positively
solved in this manuscript. We derive a new class of solution for the equations
of fireball hydrodynamics for a collectively expanding and rotating, strongly
interacting perfect fluid that contains a multi-component hadronic matter.  We
also obtain the slope parameters for various kinds of hadrons that emerge from
the collective expansion after kinetic freeze-out.

For the sake of clarity, we introduce the $i$ index to distinguish the
different type of hadrons thus the contribution of each kind of particle to the
inverse slope can be separated. Note that in this way one can find what is
really eye-catching in Fig.~\ref{fig:inverse_slope}: the inverse slope $T_i$
and as well as the single particle spectra depends on the particle type only
through the mass of the hadrons: the freeze-out temperature $T_f$ and the slope
parameter $\langle u_T\rangle$ is independent of the type $i$ of the observed
hadrons, furthermore the freeze-out temperature is a static characteristics of
the medium, apparently independent of the centrality, while the coefficient of
linearity is increasing with increasing centralities.  The result is summarized
as 
\begin{equation} 
T_i=T_f + m_i\langle u_T \rangle^2. \label{e:teffmi}
\end{equation} 
In this eq.~\eqref{e:teffmi}, the mass $m_i$ is not a smoothly
varying parameter, in contrast to eq.~\eqref{e:teffm}, but takes on the
discrete values, corresponding to the observable, PDG listed masses of those
hadron species $h_i$ that constitute the mixture in the expanding,
multi-component hadron gas.
 
\begin{figure}[h!]
	\includegraphics[scale=0.7]{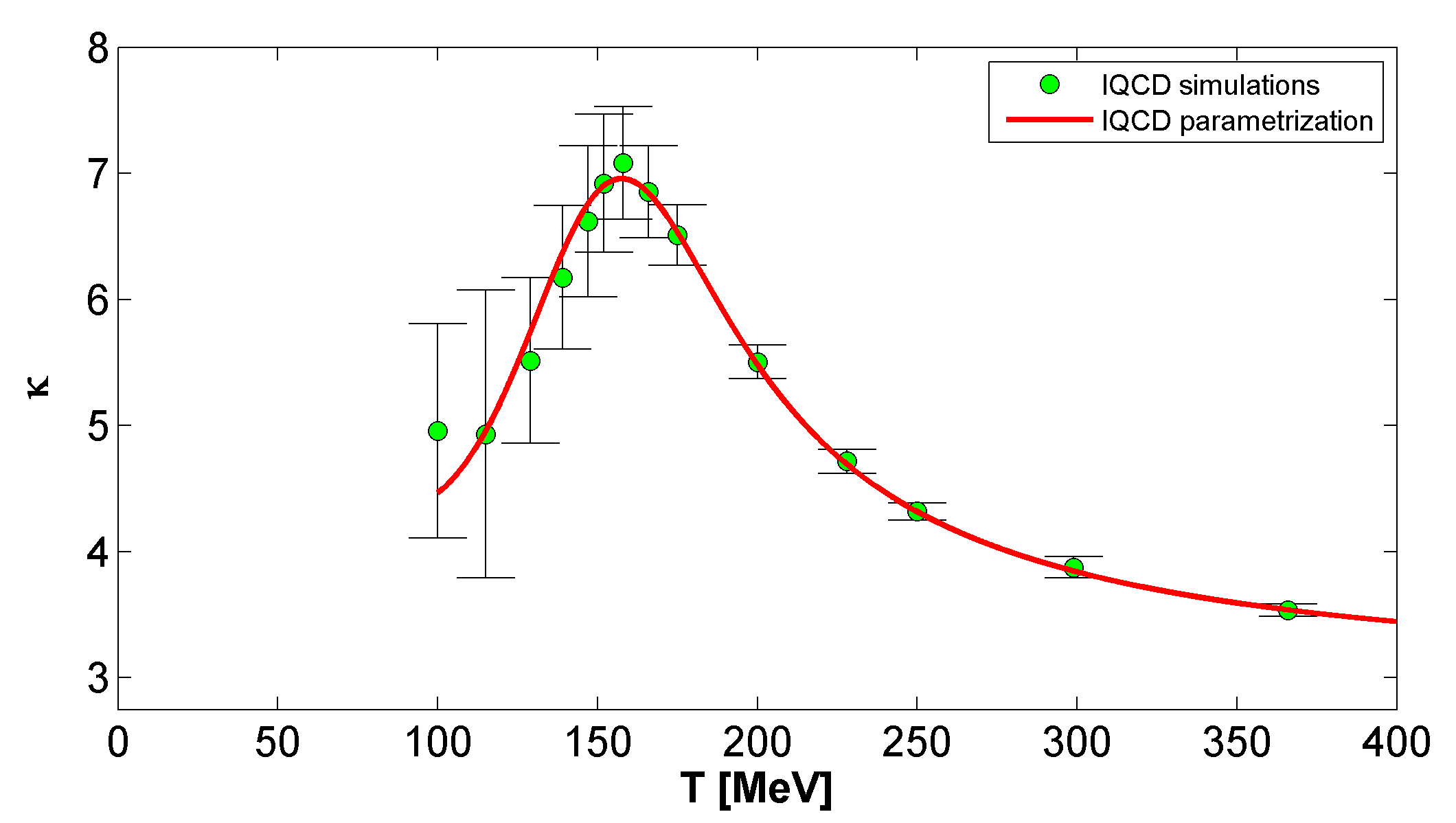}
	\centering
\caption{The result of lattice QCD simulations (green dots) on $\kappa = \varepsilon/p$ measured at $\mu_B = 0$,
         and a phenomenological parametrization (red curve), taken from ref.~\cite{Borsanyi:2010cj}.}
\label{fig:lQCD_EoS}
\end{figure}

As an extra difficulty, we also consider and are interested in describing the
time evolution of fireball hydrodynamics using a realistic, lattice QCD based
equation of state of ref.~\cite{Borsanyi:2010cj}. This excercise is a kind of
academic part of our study, motivated by the fact that in lattice QCD
simulations, the speed of sound  and the ratio of pressure to energy density at
vanishing baryochemical potential are known to depend significantly on the
temperature \cite{Borsanyi:2010cj}.  We consider such an equation of state in
this work, indicated also on Fig.~\ref{fig:lQCD_EoS}, although in a future,
even  more realistic  scenario relativistic kinematics will have to be
considered as well.  In this sense, our present study is also preparatory, yet
difficult step, aiming at future studies using lattice QCD equation of state in
the dynamics, together with relativistic kinematics and a multi-component
hadronic matter in the final state. 

By this token, the main research topic of this manuscript is given: can the
lattice QCD Equation of State and the scaling behaviour, the mass systematics
of the slope parameter of the single particle spectra be understood in a
self-consistent, hydrodynamical picture?  


\section{Equations of relativistic as well as non-relativistic hydrodynamics}
\label{s:hydro}

The basic equations of relativistic, perfect fluid hydrodynamics consists of a
set of partial differential equations, that express the local consvervation of
entropy, momentum and energy:
\begin{equation}
\partial_{\mu}\left(\sigma u^{\mu}\right)=0,\label{e:sigma}
\end{equation}
\begin{equation}
\partial_{\nu}T^{\mu\nu}=0,\label{e:tnumu}
\end{equation}
where $\sigma$ stands for the entropy density, $T^{\mu\nu}$ is the
energy-momentum tensor of a perfect fluid in the co-moving frame:
\begin{equation}
T^{\mu \nu}=\left(\varepsilon+p\right)u^{\mu}u^{\nu} - pg^{\mu \nu}.
\end{equation}
Here the metric tensor is denoted by $g^{\mu
\nu}=\textnormal{diag}(1,-1,-1,-1)$, and $u^{\mu} = \gamma (1, \mathbf{v})$ is
the four-velocity field, normalized as $u^{\mu}u_{\mu} = 1$.  The coordinates
are denoted as $x^{\mu}=\left(t, \mathbf{r}\right) =
\left(t,r_x,r_y,r_z\right)$ and the four-momentum is defined as
$p^{\mu}=\left(E, \mathbf{p} \right) = \left(E,p_x,p_y,p_z\right)$ where the
four-momentum is on-shell,  $E=\sqrt{m^2 + \mathbf{p}^2}$.

Eq.~(\ref{e:sigma}) expresses local entropy conservation, while eq.~(\ref{e:tnumu}) expresses local conservation of energy and momentum. This equation can be projected to a component parallel to $u^{\mu}$ that yields the energy equation:
\begin{equation}
\partial_{\mu}\left(\varepsilon u^{\mu}\right) + p \partial_{\mu}u^{\mu} = 0,  \label{e:rel-energy}
\end{equation}
while the component pseudo-orthogonal to the four-velocity field
yields the relativistic Euler equation: 
\begin{equation}
\left(\varepsilon + p\right) u^{\nu }\partial_{\nu}u^{\mu}=\left(g^{\mu \nu} - u^{\mu}u^{\nu} \right)\partial_{\nu}p.
\label{e:rel-Euler}
\end{equation}
Local conservation laws of densities cannot be used generally, because in
the initial, hot temperature stage, at vanishing baryochemical potential, the
only conserved charge is the net baryon number -- which is zero in the considered case. 
As the medium cools down quark and gluon degrees of freedom get confined to hadrons. 
After hadro-chemical freeze-out, this system of equations
is supplemented with the continuity equation of particle densities. 
One new conservation law is obtained for each of the hadrochemically frozen particle types:

\begin{equation}
\partial_{\mu}\left(n_i u^{\mu}\right)=0, \qquad \mbox{\rm for} \quad {T\le T_{chem},} \quad {i=1,2,...,j}
\label{e:cont-i}
\end{equation}
where $n_i$ is the particle density of the $i^{th}$ hadron, and $j$ counts that how many kind of hadrons are frozen out hadrochemically. 

\begin{table}[hb]		
\renewcommand{\arraystretch}{2}
\begin{center}
	\begin{tabular}{|c|c|}
	\hline		
		\textbf{QM} $\left( T_0 \ge T > T_{chem} \right)$ & 
			\textbf{HM} $\left( T_{chem} \ge T \ge T_{f} \right)$ \\
	\hline
		$\varepsilon + p = T\sigma + \sum_{i=1}^j \mu_i n_i \approx T \sigma$ & 
			$\varepsilon + p = T\sigma + \sum_{i=1}^j \mu_i n_i \approx \sum_{i= 1}^j m_i n_i$ \\ 
              	$p=T\sigma /\left(1+\kappa\right)$ & 
			$p= \sum_{i= 1}^j p_i = \left(\sum_{i= 1}^j n_i \right) T $ \\ 
	\hline
	\end{tabular}
\caption{
We assume, that at high temperature, the enthalpy density is dominated by the entropy term while below
the chemical freeze-out temperature of hadrons, their rest mass dominates the same term. In these equations, $j$ counts
how many kind of different hadrons are hadrochemically frozen-out.
\label{t:EoS}
}
\end{center}	
\end{table} 

In the high temperature phase, $T > T_{chem}$, we have five equations, eqs. (\ref{e:sigma},\ref{e:rel-energy},\ref{e:rel-Euler}),
for six independent variables, the three independent spatial components of the four-velocity field $\mathbf{v}$ and
the energy, the pressure and the entropy density.
The energy equation can be transformed to a temperature equation,
given in Table~\ref{tab:relat_eqsystem}, that introduces a new variable, the temperature, 
utilizing also the $\varepsilon=\kappa(T) p$ equation of state, resulting in five equations
for six independent quantities, but replacing the unknown energy density with the unknown temperature field and a known $\kappa(T)$ function.
This set of equations becomes closed with the fundamental equation of thermodynamics, 
and the assumption that the enthalpy is dominated by entropy density times the temperature at such a high temperatures,
summarized in the left column of Table~\ref{t:EoS}.  

The state before the hadronization corresponds to the quark matter (QM) and
after the transition the hadronic matter (HM) is formed. Different
thermodynamical approaches may be used to describe the different phases. The
general form of the fundamental equation of thermodynamics is:

\begin{equation}
\varepsilon+p=T\sigma + \sum\limits_i \mu_i n_i.
\end{equation}
In heavy ion collisions at RHIC and LHC energies, the bariochemical potential ($\mu$) of the strongly
coupled quark-gluon plasma is close to zero. Consequently in this case we can
use the
\begin{equation}
\varepsilon+p\approx T\sigma
\end{equation}
approximation. As the medium cools down and a chemically frozen hadron gas is
formed, the hadrochemical potential approaches the particle mass $m_i$ for
hadron type $i$ and the $T\sigma$ term becomes negligible compared to the
energy of the rest-mass. In this way, in the chemically frozen hadronic matter one may
approximate the enthalpy density as
\begin{equation}
\varepsilon+p\approx \sum\limits_i m_i n_i.
\end{equation}

At lower temperatures, $T \le T_{chem}$, we have additional relations, namely the continuity equations for each of the chemically frozen hadronic
species. Suppose that there are $j$ such chemically frozen hadrons ($i = 1, ..., j$), then we have $j-1$ independent new equations, in addition to the
entropy conservation. This way we have $j+4$ equations for $j+5$ unknowns, that we take as the $j$ different densities, the temperature, the pressure,
and the three independent spatial components of the four-velocity field.
Similarly to the high temperature case, the energy equation can be transformed again to a temperature equation,
given in the right column of Table~\ref{tab:relat_eqsystem}. Indeed, this transformation
utilizes also the $\varepsilon=\kappa(T) p$ equation of state, similarly to the high temperature case, and results again in $j+4$ equations
for $j+5$ unknowns. The $\kappa(T)$ function is assumed to be known from lattice QCD.
This set of equations is closed, similarly to the high temperature phase, with the fundamental equation of thermodynamics, 
and the assumption that the chemical potentials are given approximately by the masses of hadrons and
that the sum over chemical potentials times the corresponding densities is the dominant term in this low temperature phase,
as summarized in the right column of Table~\ref{t:EoS}. Actually, such an approximation is justified in the non-relativistic
kinematic region, as it leads naturally to the usual form of the non-relativistic Euler equation.

\begin{table}[ht]
\renewcommand{\arraystretch}{2}
\begin{center}
	\begin{tabular}{|c|c|}
	\hline		
		\textbf{QM} $\left( T_0 \ge T > T_{chem} \right)$ & 
			\textbf{HM} $\left( T_{chem} \ge T \ge T_{f} \right)$ \\
	\hline
		$\partial_{\mu}\left(\sigma u^{\mu} \right) = 0$ & 
			$\partial_{\mu}\left(n_i u^{\mu} \right) = 0,\:\:\: i = 1, ... , j$ \\
		$\left[ \frac{1+\kappa}{T}\frac{d}{dT}\left(\frac{T\kappa}{1+\kappa}\right)\right]u_{\mu}\partial^{\mu} T + \partial_{\mu} u^{\mu}=0$ & 
			$\left[ \frac{1}{T}\frac{d}{dT}\left(\kappa T \right)\right]u_{\mu}\partial^{\mu} T + \partial_{\mu} u^{\mu}=0$ \\ 
		$\left[ \sigma  T\right] 
			u^{\nu }\partial_{\nu}u^{\mu}=\left(g^{\mu \nu} - u^{\mu}u^{\nu} \right)\partial_{\nu}
				\left[\frac{\sigma T}{1+\kappa}\right] $  & 
			$\left[\sum_{i= 1}^j  m_i n_i\right] 
				u^{\nu }\partial_{\nu}u^{\mu}=\left[g^{\mu \nu} - u^{\mu}u^{\nu} \right]\partial_{\nu} 
				\left[ \sum_{i=1}^j n_i T\right] $ \\
	\hline 
	\end{tabular}
\caption{
	The closed system of partial differential equations, corresponding to relativistic hydrodynamics,  
	as specified for quark matter (QM) and for a chemically frozen, multi-component hadronic matter (HM). 
	Utilizing the equations of state from Table~\ref{t:EoS}, we obtain 5 equations for 5 unknowns in the high temperature, QM phase,
	while we obtain $j+4$ equations for $j+4$ unknowns in the chemically frozen, multi-component HM phase, where $j$ counts the number of
	type of chemically frozen hadrons. \label{tab:relat_eqsystem}
}
\end{center}	
\end{table}

\begin{table}[hb]	
\renewcommand{\arraystretch}{2}
	\begin{center}
	\begin{tabular}{|c|c|}
	\hline		
		\textbf{QM} $\left( T_0 \ge T > T_{chem} \right)$ 
		& \textbf{HM} $\left( T_{chem} \ge T \ge T_{f} \right)$ \\
	\hline
		$\partial_t \sigma +\mathbf{\nabla}\left(\sigma \mathbf{v}\right)=0$ & 
			$\partial_t n_i +\mathbf{\nabla}\left(n_i\mathbf{v}\right)=0,\:\:\: i = 1, ... ,  j$ \\
		$\left[\frac{1+\kappa}{T}\frac{d}{dT} \left(\frac {T \kappa}{1+\kappa} \right)\right]
		\left(\partial_t+\mathbf{v}\mathbf{\nabla}\right)T +\mathbf{\nabla} \mathbf{v}=0$ & 
			$\left[\frac{1}{T} \frac{d}{dT} \left(\kappa T \right)\right]\left(\partial_t+\mathbf{v}\mathbf{\nabla}\right)T+\mathbf{\nabla} \mathbf{v} =0$\\ 
		$\left[ \sigma  T\right] 
		       \left(\partial_t+\mathbf{v}\mathbf{\nabla}\right)\mathbf{v}=-\mathbf{\nabla} 
				\left[\frac{\sigma T}{1+\kappa}\right] $  & 
			$\left[\sum_{i= 1}^j  m_i n_i\right] 
			\left(\partial_t+ \mathbf{v}\mathbf{\nabla} \right)\mathbf{v}=- \mathbf{\nabla} \left[ \sum\limits_{i=1}^j n_i T \right]$ \\ 
	\hline
	\end{tabular}
\caption{
	The system of partial differential equations that describe non-relativistic
	fireball hydrodynamics of quark matter (QM) at high, $T > T_{chem}$ temperatures and a hadrochemically frozen, multi-component hadronic
	matter (HM) at lower, $T \le T_{chem}$ temperatures.
	Similarly to the relativistic case, we have obtained 5 equations for 5 unknowns in the high temperature, QM phase,
	while we have $j+4$ equations for $j+4$ unknowns in the chemically frozen, multi-component HM phase. 
	\label{tab:nonrelat_eqsystem}
}
	\end{center}	
\end{table}

The medium behaves differently before and after the hadronization which is
manifested in the different system of equations in the corresponding hydrodynamical description. The relativistic
system of equations of the two phases are summarized in Table~\ref{tab:relat_eqsystem}.

From the energy equation, we obtain a differential equation for the
temperature ($T$) by utilizing the equation of
state and the expressions for the pressure. 
Note that this procedure can be followed not only in the case of relativistic kinematics, as
summarized in Table~\ref{tab:relat_eqsystem}, but also exactly the same
method can be used to obtain the temperature equation in case of non-relativistic kinematics,
and this way one obtains a striking similarity between the system of partial differential equations
of fireball hydrodynamics both in the relativistic, and in the non-relativistic kinematic domain.
Consequently in the $|\mathbf{v}|^2 \ll 1$ approximation, the system of partial differential equations of relativistic hydrodynamics of Table~\ref{tab:relat_eqsystem} 
directly correspond
to the system of partial differential equations of non-relativistic hydrodynamics,
as summarized in Table~\ref{tab:nonrelat_eqsystem}.

\section{Exact and analytic solutions of fireball hydrodynamics}
\label{s:rotsol}

In this section, we focus on the non-relativistic approximation, for the sake
of clarity and simplicity, leaving the discussion of the relativistic kinematics to a
follow-up publication. Although such a non-relativistic approximation limits the direct
applicability of our results to a detailed comparision with particle
production to low transverse momentum and to mid-rapidity, similar  
simple and exact non-relativistic fireball solutions provided already important insights
to the mass systematics of the single-particle spectra at low $p_T$ and at
mid-rapidity. 

Linear mass dependencies of the slope parameters of the single particle spectra
were obtained before in exact solutions of non-relativistic fireball
hydrodynamics, where a spatially homogeneous temperature profile was matched
with a Gaussian density profile, both for spherically~\cite{Csizmadia:1998ef}
and for ellipsoidally symmetrically expanding
fireballs~\cite{Akkelin:2000ex,Csorgo:2001xm}, corresponding to central and
non-central heavy ion collisions, respectively. However, these Gaussian
solutions are readily generalized to an arbitrary but still spherically~\cite{Csorgo:1998yk} or ellipsoidally symmetric temperature profiles~\cite{Csorgo:2001ru}, where maxima of the temperature profile function correspond to local minima in the matching density profile functions.
Similarly, axially or ellipsoidally symmetric expanding fireballs are described
by certain recently found exact solutions of perfect fluid hydrodynamics in the
relativistic kinematic domain as well, that are characterized by a scaling variable 
and an arbitrary positive definite scaling function for the initial temperature profile,
that is matched with a corresponding density profile~\cite{Csorgo:2003rt,Csorgo:2003ry}.

We have summarized in
Tables~\ref{t:EoS},\ref{tab:relat_eqsystem},\ref{tab:nonrelat_eqsystem} the
systems of partial differential equations of a fireball that is created in
(non-)relativistic heavy-ion collisions. It is important to note that in the
temperature equation the coefficient of the co-moving derivative of the
temperature is the same in both Tables~\ref{tab:relat_eqsystem} and
\ref{tab:nonrelat_eqsystem}, and this similarity between the relativistic and
non-relativistic kinematics may play a role in subsequent, future studies.

Given such a background, let us present in this section two new, rotating exact
solutions of non-relativistic fireball hydrodynamics. Their validity can be straigthforwardly
tested with the help of Table~\ref{tab:nonrelat_eqsystem},
so we do not detail their derivation here. These solutions listed in the following subsections
are  given in the laboratory frame $K$, assuming a collider type of experiment so that the center of the
fireball is at rest in the laboratory frame $K$.

The exact and analytic solutions that are already known, 
can be grouped in a new manner as well, into three different kind of solutions, noted
first in ref.~\cite{Csorgo:2013ksa}.

\begin{enumerate}
\item The first kind of scaling solutions of fireball hydrodynamics are characterized by spatially homogeneous temperature profile and
corresponding Gaussian (entropy) density profile. These solutions
may also feature a realistic, temperature dependent energy density/pressure ratio,
or a corresponding temperature dependent speed of sound, see
 for example refs.
\cite{Csizmadia:1998ef,Akkelin:2000ex,Csorgo:2001xm,Csorgo:2002kt,Csorgo:2013ksa,Nagy:2016uiz}.
\item The second kind of scaling solutions of fireball hydrodynamics
are characterized by arbitrary, spatially inhomogeneous temperature profiles
and  corresponding, matching density profiles. The price for having the freedom of  
a realistic, arbitrary initial temperature profile 
is the need to have a temperature independent  speed of sound,
see refs.~\cite{Csorgo:1998yk,Csorgo:2001ru,Csorgo:2003rt,Csorgo:2003ry,Nagy:2009eq,Csanad:2012hr,Csorgo:2013ksa}.
\item 
The existence of third  kind of scaling solutions of fireball hydrodynamics was
noted first in ref.~\cite{Csorgo:2013ksa}, but elaborated only in ref.~\cite{Csorgo:2016ypf}.  These third kind of solutions may gain further
importance in the future,  as they allow to use a lattice QCD equation of state
in a realistic, spatially inhomogeneous, scaling solution of fireball
hydrodynamics.
\end{enumerate}

\subsection{New solutions of the first kind:\\
Exact, parametric solutions of rotating and expanding fireballs with Gaussian density profiles}

The first  kind of scaling solutions of fireball hydrodynamics correspond to
homogeneous temperature profiles, 
with a corresponding Gaussian (entropy)density profile,  
and a temperature
dependent $\kappa = \kappa(T) = \varepsilon/p$ function.

\begin{table}[ht]
\renewcommand{\arraystretch}{1.8}
\begin{center}
\begin{tabular}{|rrr|}
\hline
\multicolumn{3}{|c|}{\bf Relations valid both in Quark Matter (QM) and in Hadronic Matter (HM):}  \\\hline
\multicolumn{3}{|c|}{ 
		$V = (2 \pi)^{3/2} XYZ$,
		}  \\
\multicolumn{3}{|c|}{
 $X\big(\ddot X-\omega^2R\big) = Y\ddot Y = Z\big(\ddot Z -\omega^2R\big) = D$,
  } \\ [0.05ex] \hline 
\multicolumn{3}{|c|}{
\null \hfill	
$\dot\vartheta\equiv\fracd{\omega}{2}$, 
 \hfill
$\omega=\omega_0\fracd{R_0^2}{R^2}$, 
 \hfill
$R = \fracd{X+Z}{2}$, 
 \hfill \null
       } 
\\ 
\multicolumn{3}{|c|}{
\null \hfill	
$H_x=\fracd{\dot X}{X}$, 
 \hfill
$H_y=\fracd{\dot Y}{Y}$, 
 \hfill
$H_z=\fracd{\dot Z}{Z}$,
 \hfill \null
       } 
\\ [0.05ex]
\hline 
\multicolumn{3}{|c|}{ $s = \fracd{r_x^2}{X^2}+\fracd{r_y^2}{Y^2}+\fracd{r_z^2}{Z^2} +
	 \z{\recd{Z^2}-\recd{X^2}}\sz{(r_x^2-r_z^2)\sin^2\vartheta+r_xr_z\sin(2\vartheta)} $, } \\[0.05ex] \hline 
\multicolumn{3}{|c|}{ 
		$ \v v\z{\v r,t} = \v v_H\z{\v r,t} + \v v_R\z{\v r,t} $,
		}  \\
\multicolumn{3}{|c|}{
$ \v v_H\z{\v r,t} = \begin{pmatrix} (H_x\m{cos}^2\vartheta+H_z\m{sin}^2\vartheta)r_x \\ H_yr_y \\ (H_x\m{sin}^2\vartheta+H_z\m{cos}^2\vartheta)r_z
                     \end{pmatrix} + (H_z-H_x)\fracd{\sin(2\vartheta)}{2}\begin{pmatrix} r_z\\0\\r_x \end{pmatrix} $ , } \\
\multicolumn{3}{|c|}{
$ \v v_R\z{\v r,t} = 
                      \dot\vartheta \begin{pmatrix} r_z \\0\\ - r_x \end{pmatrix}
                     + \dot\vartheta \begin{pmatrix} \z{\fracd{X}{Z}\m{cos}^2\vartheta+ \fracd{Z}{X} \m{sin}^2\vartheta} r_z \\0\\ -\z{\fracd{X}{Z}\m{sin}^2\vartheta
                                    +\fracd{Z}{X} \m{cos}^2\vartheta} r_x \end{pmatrix}
                     + \dot\vartheta \left(\fracd{X}{Z} - \fracd{Z}{X}\right) \fracd{\sin(2\vartheta)}{2}\begin{pmatrix} r_x\\0\\-r_z \end{pmatrix} $,
  } \\  
\hline 
\end{tabular}
\end{center}
\caption{
Equation of state {\it independent} part of new, exact,  rotating solutions of fireball hydrodynamics,
as given in the laboratory frame $K$, where a rotating ellipsoid with
time dependent principal axis $X$, $Y$ and $Z$ is not only expanding but also spinning around the $r_y$ direction. 
The time dependent parameter $\vartheta$ stands for the
angle between the principal axis $X$ and the impact parameter direction, $r_x$ in the intertia frame $K$. 
The first of this kind of solutions was given in  
ref.~\cite{Nagy:2016uiz}, 
that we generalize in this work to quark matter and multi-component hadronic matter scenarios.
}
\label{t:valid-both-QMHM}
\end{table}

In this subsection we generalize the results of ref.~\cite{Nagy:2016uiz}, 
to a quark matter rehadronizing to a multi-component hadron gas.
Following ref.~\cite{Nagy:2016uiz}, 
we have written up the velocity field as a sum of two terms: a ,,Hubble-term'' $\v v_H$, and a ,,rotational term'' $\v v_R$. 
This form of the velocity field $\mathbf{v}$ and the scaling variable $s$ satisfies the scale equation:
\begin{eqnarray}
\left(\partial_t + \mathbf{v}\mathbf{\nabla}\right) s & = & 0,\\
\mathbf{v} & = &  \mathbf{v}_H + \mathbf{v}_{R}.
\end{eqnarray}
The Hubble term has zero curl (and thus does
not contribute to the vorticity of the flow), while the rotational term has zero divergence, hence it does not contribute
to the overall expansion of the fluid:
\begin{eqnarray}
\nabla \v v & = &  \nabla \v v_H =  \frac{\dot V}{V},\\
\nabla \v v_R & = & 0. 
\end{eqnarray}
The terms in $\v v$ that are proportional to the angular velocity $\dot\vartheta$ contribute to the rotational flow velocity $\v v_R$, which determines
the vorticity of the solution as
\begin{eqnarray}
\gvec\omega\z{\v r,t}   & = & \nabla \times\v v =  \nabla \times\v v_R  , \\
\nabla\times \v v_H & = & 0.
\end{eqnarray}
The vorticity vector is parallel with the axis of rotation, and the value of its only non-vanishing component in the laboratory frame, 
$\gvec\omega = (\omega_x,\omega_y, \omega_z)$ , where the non-vanishing
component is given as 
\begin{align}
\omega_y\z{\v r,t}   &= \omega + \frac \omega 2\z{\frac XZ +\frac ZX}.
\end{align}
The equation of state specific parts of the solution are summarized in Table~\ref{t:QM-or-HM-only}.
\begin{table}[ht]
\renewcommand{\arraystretch}{1.8}
\begin{center}
\begin{tabular}{|cc|}
\hline
  \textbf{QM} $\left( T_0 \ge T > T_{chem} \right)$  
		& \textbf{HM} $\left( T_{chem} \ge T \ge T_{f} \right)$ \\
{\textbf{Valid in QM only:} } 	& 
		{\textbf{Valid in HM only:} } 
			\\ \hline 
			 $T\equiv T(t)$, 
			 &
			 $T\equiv T(t)$,
			 \\
 $D = \fracd{1}{1+\kappa(T)},$ 
				&
		$ D = \fracd{T}{\langle m\rangle }$,
		\\
$\sigma  = \sigma_0 \fracd{V_0}{V}\exp\z{-s/2}$, 
				&
		$n_i = n_{i,0} \fracd{V_0}{V}\exp\z{-s/2}$ \quad $i = 1, ..., j$,
   \\
  $ \fracd{\dot T}{T} \sz{1+ \kappa}  \fracd{\m d }{\m dT}\sz{\fracd{T\kappa}{1+\kappa}} 
+ \fracd{\dot V}{V}=0$, 
				&
		  $ \fracd{\dot T}{T} \fracd{\m d}{\m dT} \sz{T\kappa} + \fracd{\dot V}{V}=0$. 
\\ \hline
\end{tabular}
\end{center}
\caption{
Equation of state {\it dependent} part of new, exact, rotating solutions of fireball hydrodynamics,
that completes Table~\ref{t:valid-both-QMHM}.
In this table, those parts are shown that are specific to the time evolution of quark matter or hadronic matter.
}
\label{t:QM-or-HM-only}
\end{table}
In the same table the acceleration parameter $D$ is related to $\langle m \rangle$, the average mass of the particles
in the multi-component hadronic matter:
\begin{equation}
\langle m \rangle=\frac{\sum\limits_i m_i n_i}{\sum\limits_i n_i}.
\end{equation}
In such a multi-component hadronic matter, the average mass is assumed to be time independent, $\langle m\rangle = (\sum_{i= 1}^j n_{i,0} m_i)/(\sum_{i= 1}^j n_{i,0} ) $.

In this way we reduced the complicated set of partial differential equations to
a set of ordinary differential equations. This result generalizes the triaxial,
rotating and expanding fireball solution of ref.~\cite{Nagy:2016uiz} from a
single component hadron gas, characterized by  mass $m$, to a perfect fluid of
quark matter that hadronizes to  a mixture of hadrons with average mass
$\langle m \rangle$. Table \ref{t:QM-or-HM-only} indicates that this mixture
expands and rotates together, because the scale parameters $X$, $Y$ and $Z$ as
well as the parameters of the rotation $\omega$ are independent of the index
$i$ of hadron types $h_i$.  These solutions are all (rotating) Gaussian
solutions, due to the $\exp(-s/2)$ type of terms in the entropy and chemically
frozen density profiles given in Table~\ref{t:QM-or-HM-only}, that correspond
to a position independent, but time dependent temperature profiles.

\subsection{Fireball explosion at the QCD Critical Point}

The last line of Table~\ref{t:QM-or-HM-only}  indicates a beautiful exact
and analytic result: Although the lattice QCD EoS suggests that the function
$\kappa(T)$ is a smoothly varying function of the temperature as indicated on
Fig.~\ref{fig:lQCD_EoS}, the coefficient of ${\dot T}/{T}$ in the last line
of Table~\ref{t:QM-or-HM-only} actually changes. The difference of these coefficients is
\begin{equation}
  \sz{1+ \kappa}  \fracd{\m d }{\m dT}\sz{\fracd{T\kappa}{1+\kappa}} 
		  - \fracd{\m d}{\m dT} \sz{T\kappa} = 
		- \fracd{T \kappa}{1+\kappa} \fracd{{\m d}\kappa}{{\m d}T},
\end{equation}
which vanishes at the temperature of $T_{max}\approx 158 $ MeV, where the $\kappa(T)$
function has a maximum and changes sign at this point: 
\begin{itemize}
\item{ If $T_{chem} > T_{max}$,} 
then the coefficient of ${\dot T}/T$ increases, and the rate of
change of the temperature decreases at the same logarithmic derivative of the
volume ${\dot V}/V$, characterizing the volume expansion at hadrochemical
freeze-out. 
\item{If $T_{chem} = T_{max}$,} the time evolution of the
temperature changes smoothly at the hadrochemical freeze-out.
\item{If $T_{chem} < T_{max}$,} 
the coefficient of ${\dot T}/T$ decreases, and the rate of
change of the temperature increases at the same logarithmic derivative of the
volume ${\dot V}/V$.
\end{itemize}
As we shall see below, not only the rate of change of the temperature with
the increase of the volume is sensitive to quark confinement and subsequent
 hadrochemical freeze-out, but also the dimensionless measure of the rate of
acceleration of the expansion changes at the chemical freeze-out, due to the
changes in the expansion dynamics.

We assume that the hadronization and the hadrochemical freeze-out happens
nearly simultaneously, as a consequence $T_{chem}\approx T_c$ in this case. 
For a spatially homogeneous temperature profile, and a simultaneous
quark confinement and hadrochemical freeze-out, the boundary conditions are:
\begin{align}
T_Q(t_{chem})&=T_H(t_{chem})=T_{chem},\\
\mathbf{v}_Q(\mathbf{r},t_{chem})&=\mathbf{v}_H(\mathbf{r},t_{chem}),\\
\kappa_Q(T_{chem})&=\kappa_H(T_{chem}).
\end{align}
Due to these special boundary conditions one can obtain an important 
relation for $D$ that is a dimensionless measure of the
rate of expansions of the scales in the equations of motion:
\begin{equation}\label{eq:second_exp}
D_Q(T_{chem}) = \frac{1}{1+\kappa_{chem}}\approx 0.13 < D_H (T_{chem}) = \frac{T_{chem}}{\langle m\rangle} \approx 0.59.
\end{equation}
The expression of the left side in eq. \eqref{eq:second_exp} characterizes the
dimensionless acceleration parameter $D_Q$ in the quark phase, and the right side of
the same relation relates $D_H$ to the acceleration of the hadronic matter. 
Given that $D_H > D_Q$,
the acceleration of the scales jumps at $t_{chem}$ in every direction,
so the directional Hubble constants $\dot X/X$, $\dot Y / Y$ and $\dot Z/Z$
have a break at these temperatures, and each of the scales $X$, $Y$ and $Z$ starts
to expand faster as the temperature cools below the hadrochemical freeze-out
temperature $T_{chem}$.  We propose to call this
phenomenon as   a second explosion in the Little Bangs or heavy ion collisions.
The first explosion is the violent expansion that starts just after collision,
due to large initial energy densities, temperatures and pressure gradients,
while this second explosion starts just after the
conversion to the hadronic phase and appears
due to the role of hadrochemical freeze-out.
 The effect of the second explosion becomes 
smoothened, if the average mass in the fireball is increased at $T\approx
T_{chem}$. We expect, but cannot detail in this paper that a smooth cross-over
may further slow down the expansion dynamics,
but the detailed discussion of such a phenomena goes beyond the scope 
of this manuscript.

Our calculations seem to be laying the ground for similar calculations in the
relativistic kinematic domain. Before looking for such  new solutions of
relativistic hydrodynamics, let us perform a consistency check and see if the
single particle spectra from such solutions looks to be realistic or not.

\section{Evaluation of the single-particle spectra}

The single particle spectra has already been calculated from single-component (SC)
hydrodynamical solutions \cite{Csorgo:2015scx, Nagy:2016uiz}.
In this section we follow the notations and conventions of these earlier works,
without giving the full details,  referring also  to the Introduction
for a motivation, and highlighting only the main, characteristic  
features of our results.

The slope parameters of the single particle spectra are apparently simple 
also in our multi-component (MC) generalization of rotating and expanding
fireballs, when given  in the $K'$ frame that co-rotates with the fireball:
\begin{equation}
\label{momdist_in_K'}
N_{1,i}(p'_i) \propto \exp\{-\frac{1}{2m}\sum_{k,l=1}^3 p'_{k,i}\left(\textbf{T}'\right)^{-1}_{kl,i}p'_{l,i}\},
\end{equation}
where $k,l=\{1,2,3\}=\{x,y,z\}$ and 
$\left(\textbf{T}'\right)^{-1}_{kl,i}$ is 
the inverse of the effective temperature matrix or in other words the inverse slope
matrix. 

Similarly to the hydrodynamical solutions of a single-component hadron
gas~\cite{Csorgo:2015scx, Nagy:2016uiz}, this single particle spectra can
be transferred to the laboratory frame and we obtain:
\begin{equation}
	N_{1,i}(p_i) \propto \exp\{-\frac{1}{2m}\sum_{k,l=1}^3 p_{k,i}\left(\textbf{T}\right)^{-1}_{kl,i}p_{l,i}\}.
\end{equation}

We have collected the matrix elements for multi-component (MC) and single
component (SC) hadronic matter (HM) in Table~\ref{tab:inverse_slopes}, where
we denote these final state hadronic observables with the subscript $f$ that
denotes kinetic freeze-out. Note that for the non-rotating fireballs, the
inverse temperature matrix is diagonal in the center of mass frame $K$, while
for rotating fireballs, the inverse temperature matrix is diagonal only in the
co-rotating frame $K'$, so an additional transformation to $K$ has to be made,
similarly how it was performed in ref.~\cite{Csorgo:2015scx} .
\begin{table}[ht]
	\centering
	\renewcommand{\arraystretch}{1.5}
	\begin{tabular}{|c|c|c|}
		\hline	
		& \textbf{SC HM} \cite{Csorgo:2001xm, Csorgo:2015scx, Nagy:2016uiz} 
			& \textbf{MC HM} \\
		\hline
		& $T_x=T_f+m \dot{X_f}^2$ & $T_{x,i}=T_f+m_i \dot{X_f}^2$ \\
		$\omega_0=0$ & $T_y=T_f+m \dot{Y_f}^2$ & $T_{y,i}=T_f+m_i \dot{Y_f}^2$ \\
		& $T_z=T_f+m \dot{Z_f}^2$ & $T_{z,i}=T_f+m_i \dot{Z_f}^2$\\ \hline
		& $T'_{xx}=T_f+m\left(\dot{X_f}^2+\omega_f^2 R_f^2\right)$ & $T'_{xx,i}=T_f+m_i\left(\dot{X_f}^2+\omega_f^2 R_f^2\right)$ \\
		$\omega_0\neq0$ & $T'_{yy}=T_f+m \dot{Y_f}^2$ & $T'_{yy,i}=T_f+m_i \dot{Y_f}^2$ \\
	($K'$ frame)& 
	$T'_{zz}= T_f+m\left(\dot{Z_f}^2+\omega_f^2 R_f^2\right)$ 
	& $T'_{zz,i}=T_f+m_i\left(\dot{Z_f}^2+\omega_f^2 R_f^2\right)$ 
		\\ \hline			
	\end{tabular}
	\caption{The comparison of the inverse slope parameters after
 	kinetic freeze-out at temperature $T_f$,
	for a Single Component (SC)  and for a  Multi-Component  (MC) 
	Hadronic Matter (HM) from exact and analytic solutions
	of fireball hydrodynamics, in the case of rotating and nonrotating 
	ellipsoid as well. In case of non-rotating fireballs,
	the temperature matrix is diagonal in the center of mass frame $K$,
	$T_{kl}=\textnormal{diag}(T_{x},T_{y},T_{z})$. 
	For rotating fireballs,
	the temperature matrix is diagonal only in the co-rotating frame
	$K^{'}$, where $T^{'}_{kl} = \textnormal{diag}
	(T_{xx}^{'},T_{yy}^{'},T_{zz}^{'})$.
	}	
	\label{tab:inverse_slopes}
\end{table}

We thus find that starting from a high temperature quark matter phase, followed
by a nearly simultaneous quark confinement and hadrochemical freeze-out, we
obtain apparently thermal single-particle spectra, where the inverse slope
parameters have linear mass for each of the hadronic components $h_i$ and they
depend on the particle types only through their masses $m_i$, similarly to the
experimental data summarized in the Introduction.
After kinetic freeze-out, the effective temperature $T_{i}$ of
the single-particle spectra of hadron type $h_i$  is found to be a sum of the
kinetic freeze-out temperature $T_f$ (that is independent of the hadron type
$h_i$) and  a term proportional to the mass $m_i$ of hadron type $h_i$.  
The coefficient of proportionality to $m_i$ is also found to be 
independent of the hadron type $h_i$ but be dependent on the radial flow (and 
vorticity) of collective dynamics.

As a conclusion, the presented new class of multi-component, exact solution of
fireball hydrodynamics provides a clear-cut, but not-yet-relativistic
explanation for the mass-scaling behaviour of the single particle spectra in
heavy ion collisions.  Scaling violations, corresponding 
to deviations from the linear mass scaling may be expected
due to relativistic kinematics and temperature gradients, as discussed in
ref.~\cite{Csorgo:1995bi}, or, due to possible non-equilibrium kinetic
freeze-out effects~\cite{Sinyukov:2002if}.

\section{Summary}

We have presented new, exact, parametric solutions of fireball hydrodynamics
for a quark matter hadronizing to a chemically frozen, multi-component
hadronic matter. These solutions have spatially homogeneous initial temperature
profiles and  corresponding Gaussian (entropy) density profiles.
We have identified a change in the time evolution of the temperature
that is due to the quark confinement to hadrons and to the simultaneous
hadrochemical freeze-out.
In addition, we have found a new phenomenon at the
hadrochemical freeze-out, that relates to the dynamics of the hadronizing
fireball: a second explosion, that seems to be, rather counter-intuitively, the
strongest if the energy density to pressure ratio is the largest (frequently
referred to as the softest point of QCD).  This second explosion is caused
by the increase of the acceleration after hadrochemical freeze-out, which
effect seems to appear due to the change of the thermodynamics from a quark
matter where no conserved charges exist to a chemically frozen hadronic matter,
where each of the chemically frozen hadronic types obeys a local continuity
equation.  We conjecture, that this second explosion is washed out if the
transition from quark matter to hadronic matter corresponds to a continuous crossover,
however, we could not detail this conjecture and elaborate the 
description of such a cross-over transition in this manuscript.  

We have successfully generalized earlier exact and analytic hydrodynamical
solutions from a single-component to a multi-component hadronic matter and we
have shown that after the hadrochemical  freeze-out of the multi-component
hadronic matter, the slope parameters of the single particle spectra for
various hadrons follow the experimentally observed affine-linear mass-scaling
behaviour.  Scaling violations, or deviations from such an affine-linear
mass-scaling  of the slope parameters may also be expected, due to relativistic
kinematics and/or temperature gradients~\cite{Csorgo:1995bi}, or, further
possible non-equilibrium effects~\cite{Sinyukov:2002if}.

\section*{Acknowledgments}
T. Cs. would like to thank to D. Anchiskin, L.P. Csernai, Y. Hatta, D. Klabucar,
T. Kunihiro, S. L\"ok\"os, K. Ozawa, P. Petreczky and Yu. M. Sinyukov
for inspiring discussions and to K. Ozawa for his kind hospitality at KEK, Tsukuba, Japan.
This work was partially  supported by a KEK Visitor Fund (Japan), 
and by the EFOP 3.6.1-16-2016-00001, the OTKA NK
101438 and the NKTIH FK 123842 FK 123959 grants (Hungary) as well as by  the
bilateral exchange programme of the Hungarian and the Ukrainian Academies of Sciences,
grants NKM-82/2016 and NKM-92/2017.



\begin{thebibliography}{999}

\bibitem{Bearden:1996dd} 
I.~G. Bearden et~al [NA44 Collaboration],
\newblock 
\newblock {\em Phys. Rev. Lett.}, {\bf 78}, 2080--2083 (1997)

\bibitem{Adler:2003cb}
S.~S. Adler et~al [PHENIX Collaboration],
\newblock 
\newblock {\em Phys. Rev.}, C {\bf 69}, 034909 (2004)

\bibitem{Schnedermann:1993ws} 
  E.~Schnedermann, J.~Sollfrank and U.~W.~Heinz,
 \newblock  {\em Phys.\ Rev.}\ C {\bf 48}, 2462 (1993)

\bibitem{Csorgo:1994fg} 
  T.~Cs\"org\H{o}, B.~L\"orstad and J.~Zim\'anyi,
\newblock   {\em Phys.\ Lett.}\ B {\bf 338}, 134 (1994)

\bibitem{Csorgo:1995bi} 
  T.~Cs\"org\H{o} and B.~L\"orstad,
\newblock   {\em Phys.\ Rev.}\ C {\bf 54}, 1390 (1996)

\bibitem{Csizmadia:1998ef}
P.~Csizmadia, T.~Cs\"org\H{o}, and B.~Luk\'acs,
\newblock 
\newblock {\em Phys. Lett.}, B {\bf 443}, 21--25 (1998)


\bibitem{Akkelin:2000ex}
S.V. Akkelin, T.~Cs\"org\H{o}, B.~Luk\'acs, Yu.M. Sinyukov, and M.~Weiner,
\newblock 
\newblock {\em Phys. Lett.}, B {\bf 505}, 64--70 (2001)

\bibitem{Csorgo:2001xm}
T.~Cs\"org\H{o}, S.V. Akkelin, Y.~Hama, B.~Luk\'acs, and Yu.M. Sinyukov,
\newblock 
\newblock {\em Phys. Rev.}, C {\bf 67}, 034904 (2003)

\bibitem{Bondorf:1978kz}
J.~P. Bondorf, S.~I.~A. Garpman, and J.~Zim\'anyi,
\newblock 
\newblock {\em Nucl. Phys.}, A {\bf 296}, 320--332 (1978)

\bibitem{Borsanyi:2010cj}
Sz. Bors\'anyi, G.~Endr\H{o}di, Z.~Fodor, A.~Jakov\'ac, S.~D. Katz, et~al.
\newblock 
\newblock {\em JHEP}, {\bf 1011}, 077 (2010)

\bibitem{Csorgo:1998yk}
T.~Cs\"org\H{o},
\newblock 
\newblock {\em Central Eur. J. Phys.}, {\bf 2}, 556--565 (2004)

\bibitem{Csorgo:2001ru}
T.~Cs\"org\H{o},
\newblock 
\newblock {\em Acta Phys. Polon.}, B {\bf 37}, 483--494 (2006)

\bibitem{Csorgo:2002kt} 
  T.~Cs\"org\H{o} and J.~Zim\'anyi,
\newblock {\em  Acta Phys.\ Hung.}\ A {\bf 17}, 281 (2003)

\bibitem{Csorgo:2003rt} 
  T.~Cs\"org\H{o}, F.~Grassi, Y.~Hama and T.~Kodama,
\newblock  {\em Phys.\ Lett.}\ B {\bf 565}, 107 (2003)

\bibitem{Csorgo:2003ry} 
  T.~Cs\"org\H{o}, L.~P.~Csernai, Y.~Hama and T.~Kodama,
\newblock {\em  Acta Phys.\ Hung.}\ A {\bf 21}, 73 (2004)


\bibitem{Nagy:2009eq} 
  M.~I.~Nagy,
\newblock   {\em Phys.\ Rev.}\ C {\bf 83}, 054901 (2011)



\bibitem{Csanad:2012hr} 
  M.~Csan\'ad, M.~I.~Nagy and S.~L\"ok\"os,
\newblock {\em  Eur.\ Phys.\ J.}\ A {\bf 48}, 173 (2012)
  [arXiv:1205.5965 [nucl-th]].

\bibitem{Csorgo:2013ksa}
T.~Cs\"org\H{o}, and M.I. Nagy,
\newblock 
\newblock {\em Phys.Rev.}, C {\bf 89}, 044901 (2014)

\bibitem{Csorgo:2015scx}
T.~Cs\"org\H{o}, M.~I. Nagy, and I.~F. Barna,
\newblock 
\newblock {\em Phys. Rev.}, C {\bf 93}, 024916 (2016)

\bibitem{Nagy:2016uiz}
M.~I.~Nagy and T.~Cs\"org\H{o},
\newblock {\em Phys.\ Rev.}\ C {\bf 94}, 064906 (2016)


\bibitem{Csorgo:2016ypf} 
  T.~Cs\"org\H{o} and G.~Kasza,
\newblock {\em JCEGI} {\bf 5}, pp. 19-32 (2017),  
	[arXiv:1610.02197 [nucl-th]].


\bibitem{KaszaMSc}
G.~Kasza,
\newblock {\textit{MSc Thesis}, E\"otv\"os University, 2017 (in Hungarian)}.

\bibitem{Sinyukov:2002if} 
  Y.~M.~Sinyukov, S.~V.~Akkelin and Y.~Hama,
\newblock {\em  Phys.\ Rev.\ Lett.}\  {\bf 89}, 052301 (2002)


\end{thebibliography}
\end{document}